\input{psfig.sty}
\documentclass[11pt,a4paper]{article}
\usepackage{jcappub}
\usepackage{array,multirow,graphicx,amssymb}
\usepackage{longtable}
\usepackage{supertabular,booktabs}
\usepackage{multicol}
\usepackage[utf8]{inputenc}

\title{Chandrasekhar's Dynamical Friction and non-extensive statistics}
\author[a]{J. M. Silva,}
\author[b]{J. A. S. Lima,}
\author[b]{R. E. de Souza,}
\author[c,d,e]{A. Del Popolo,}
\author[f]{Morgan Le~Delliou,}
\author[g]{Xi-Guo Lee}
\affiliation[a]{Centro de Forma\c{c}\~ao de Professores da UFCG, Cajazeiras, PB}
\affiliation[b]{Instituto de Astronomia, Geof\'{\i}sica e Ci\^encias Atmosf\'ericas, USP, 05508-900 S\~ao Paulo, SP, Brasil}
\affiliation[c]{Dipartimento di Fisica e Astronomia, University Of Catania, Viale Andrea Doria 6, 95125 Catania, Italy }
\affiliation[d]{INFN sezione di Catania,  Via S. Sofia 64, I-95123 Catania, Italy}
\affiliation[e]{International Institute of Physics, Universidade Federal do Rio Grande do Norte, 
59012-970 Natal, Brazil}
\affiliation[f]{Instituto de Física Teorica, Universidade Estadual de São Paulo (IFT-UNESP),\\
 Rua Dr. Bento Teobaldo Ferraz 271, Bloco 2 - Barra Funda, 01140-070
S\~{ã}o Paulo, SP Brazil}
\affiliation[g]{Institute of Modern Physics, Chinese Academy of Sciences, \\
Post Office Box 31, Lanzhou 730000, People’s Republic of China}

\abstract{
The motion of a point like object of mass $M$ passing through the background potential of massive collisionless particles ($m <<  M$) suffers a steady deceleration named dynamical friction. In his classical work, Chandrasekhar assumed a Maxwellian velocity distribution in the halo and neglected the self gravity of the wake induced by the gravitational focusing of the mass $M$. In this paper, by relaxing the validity of the Maxwellian distribution due to the presence of long range forces, we derive an analytical formula for the dynamical friction in the context of the $q$-nonextensive kinetic theory. In the extensive limiting case ($q = 1$), the classical Gaussian Chandrasekhar result is recovered. As an application, the dynamical friction timescale for Globular Clusters  spiraling to the galactic center is explicitly obtained. Our results suggest that the problem concerning the large timescale as derived by numerical $N$-body simulations or semi-analytical models can be understood as a departure from the standard extensive Maxwellian regime as measured by the Tsallis nonextensive $q$-parameter.
}

\begin{document}
\maketitle

\section{Introduction}

{

The analysis of the dynamics of stellar systems such as globular clusters or clusters of galaxies has shown that the gravitational stochastic force plays a fundamental role in their evolution (Chandrasekhar \& von Newman 1942, 1943; Kandrup 1980; Ardi \& Inagaki 1999). In these systems the stochastic force, arising from statistical fluctuations
in the number of neighbors of a test star, perturbs the stars orbits from the orbits they would have if the density distribution in the system were perfectly smooth. The existence of the stochastic force is due to the discreteness of gravitational systems, i.e. to the fact that the mass is concentrated into discrete objects like stars. The first consequence produced by the stochastic force is the existence of a frictional force that implies a preferential deceleration of a particle in the direction of motion (Chandrasekhar \& von Newman 1942, 1943). 

The study of the statistics of the fluctuating gravitational force in infinite homogeneous systems was pioneered by Chandrasekhar \& Von Neumann in two classical papers (Chandrasekhar \& Von Neumann 1942, 1943). Their analysis of the fluctuating gravitational field was formulated  statistically, in a treatment related to the so-called Holtsmark's distribution (Holtsmark 1919), by $W({\bf F})$,  which gives the probability that a test star is subject to a force F in the range ${\bf F}$+${\bf d F}$, and by the distribution $W({\bf F}$+${\bf d F}$) which gives  the speed of fluctuations, i.e. the joint probability that the star experiences a force F and a rate of change ${\bf f}$=$d {\bf F}/dt$. 
From such statistical treatment, Chandrasekhar showed the emergence of a dynamical friction (DF) force, 
%
%The concept of dynamical friction (DF) was {\it just} hinted in this statistical approach. 
%
{%\bf 
a dissipative force connected to the fluctuations in the medium,  thus an aspect of the fluctuation-dissipation relation (see also Bekenstein \& Maoz 1992).

Chandrasekhar's famous formula
%, the so called Chandresakhar's formula for 
was not obtained in the general statistical framework (Chandrasekhar \& von Newman 1942, 1943) because of the mathematical complexity of the scheme, but rather in another paper (Chandrasekhar 1943), restricted to a two-point interaction scheme, and
%
%{\it NOT NECESSARY
%As stressed by Chandrasekhar \& von Neumann (1943), for a test star moving with velocity ${\bf v}$ in a sea of field stars characterized
%by a random probability distribution of the velocities, $\Phi({\bf u})$, we may write:
%$<{\bf V}> = <{\bf u}> - {\bf v}$ where ${\bf V}$ represents the velocity of a typical field star relative to the one under consideration, ${\bf u}$
%denotes the velocity of a field star. This asymmetry of the distribution of the
%relative velocities produces, as shown by CN43, a deceleration of the test star in the
%direction of motion. This effect is dubbed as ?dynamical
%friction? (for more details see section 11 of Chandrasekhar \& von Newman 1943, ApJ 1943, 97, 1).
%}
% 
his aim to re-derive this formula only from his general theory of stochastic forces has never been realized.

In that paper (Chandrasekhar 1943), the formula was obtained assuming that a point mass moves through an infinite, homogeneous sea of field particles, in the approximation that binary encounters dominate. }
In this case a fraction of the kinetic energy of the incoming object is transferred to the stellar collisionless population whose distribution was described by a Maxwellian velocity. 

The formula shows that a massive object of mass $M$ such as a Globular Cluster passing through a background of non-colliding particles suffers a gravitational force which slows down its motion.

{%\bf 
Despite being obtained for an homogeneous and infinite system,  neglecting self-gravity (i.e., the interaction between the field particles) (Maoz 1993), and resonant interactions between the background and the infalling body (e.g., Weinberg 1986; Inoue 2009)% IN PETTS \underline{[CITE]}
, the DF mechanism is now considered a classical effect in the description and evolution of almost all many-body astrophysical systems, as it has been applied quite successfully.}

%Some examples of systems to which it has been applied 

Examples of its application involve the formation of stellar galactic nuclei via merging of old Globular Clusters (GCs) \citep{Trem1975}, the transformation from non-nucleated dwarf galaxies into nucleated ones \citep{OhL2000}, the behavior of radio galaxies in galaxy clusters \citep{Nath2008}, in nonlinear gaseous media \citep{Kin2009} and in field particles with a mass spectrum \citep{Ciot2010}.  Traditionally, such investigations were carried out in the framework of Newtonian gravity, however, alternative gravity theories like the Modified Newtonian Dynamics (MOND) has also been considered \citep{Nip2008}.

{ 
%Several authors have stressed the importance of stochastic forces and in particular 
DF is also of fundamental importance in determining the observed properties of clusters of galaxies (White 1976;
Kashlinsky 1986,1987; Colafrancesco, Antonuccio-Delogu, Del Popolo 1995) 
%while others studied the role of dynamical friction 
and in the orbital decay of a satellite moving around a galaxy or in the merging scenario (Bontekoe \& van Albada 1987; Seguin \& Dupraz 1996; Dominguez-Tenreiro \& Gomez-Flechoso 1998; Del Popolo \& Gambera 1997; Antonuccio-Delogu \& Colafrancesco 1994). 

%
%%Chandrasekhar's formula is several times misused. It is widely employed to quantify DF in a variety of situations, even if 
%%the conditions under which it was obtained are not satisfied. For example, it is used in stellar systems even if it is very %%well as in galactic systems. 
%the galaxies are not uniformly distributed (Peebles 1980; Bahcall \& Soneira 1983; Sarazin 1988; Liddle, \& Lyth 1993; White et %al. 1993; Strauss \& Willick 1995).
%
}

{%\bf  
Finally, the use of such formula speeds up considerably N-body simulations. For example, self-consistent modeling of the internal dynamics of a $10^5 M_{\odot}$ GC requires  $10^7-10^{12}$ background particles to model the inspiral: smaller mass resolution (i.e., less particles) produces an under-prediction of the dynamical friction force. The dynamical friction formula allows to skip the calculation of the interaction with the background, and to concentrate on the calculation of the internal dynamics.   
~\\

However, Chandrasekhar's formula suffers from some  break downs, such as 
\renewcommand{\theenumi}{\alph{enumi}}\begin{enumerate} 
\item the evolution of a displaced super massive black hole (Gualandris \& Merritt 2008%; IN ANTONINI
); 
\item the overprediction of the infalling timescale in cored systems, the so-called "core stalling problem" (see the following);
\item the inadequacy of the formula to describe dynamical friction in head-on encounters (Seguin \& Dupraz 1996a); 
\item the inaccuracy of the formula to calculate  DF  in disks\footnote{%\bf 
A better model to describe DF  in disks is that of obtained by Binney (1977), modifying Chandrasekhar's theory }.\end{enumerate}

It is therefore of utmost importance to have a reliable semi-analytic formula to describe DF in the break down cases. 
 
%As discussed by Petts, Gualandris \& Read (2015), 

Improving the treatment of dynamical friction was attempted, following 
two different paths: 
}
\renewcommand{\theenumi}{\Alph{enumi}}\begin{enumerate} 
{%\bf  
\item \label{enu:A}recalculate dynamical friction starting from a statistical analysis approach, whether Chandrasekhar's or another. Such approaches comprise: 
\renewcommand{\theenumii}{\arabic{enumii}}\begin{enumerate} 
{%\bf  
\item  a Fokker-Planck analysis of binary interaction to estimate the diffusion coefficients (e.g., Rosenbluth, MacDonald \& Judd 1957; Binney \& Tremaine 1987% IN Bekenstein
).  
\item the polarization cloud approach (e.g., Bekenstein \& Zamir 1991) recovers Chandrasekhar's formula in the case of very massive test particle (Kandrup 1983). 
\item the derivation of frictional effects starting from the interaction of test objects and resonant particles (e.g., Weinberg 1986). 
\item returns on Chandrasekhar's statistical approach:
\begin{enumerate} 
\item the Cohen (1975) and Kandrup (1980) two-body approximation with full stochastic theory, from which Kandrup (1983a,b) reobtained Chandrasekhar's formula for test particles more massive and slower than background particles.

Kandrup (1980) showed the stochastic approach disagree with the formula in the weak forces limit because the nontrivial role distant field stars play in the stochastic force.
\item the interaction of a test particle and a background stochastic force, going back to the Chandrasekhar's statistical theory, in Bekenstein \& Maoz (1992), and Maoz (1993), where they found Chandrasekhar's  friction force depends on the global structure of the system (Maoz 1993; Del Popolo \& Gambera 1999; Del Popolo 2003), in inhomogeneous systems, and is no longer directed opposite to the test particle's motion.

\end{enumerate}
}\end{enumerate}

}

%Binney?s model supposes that test particles have an anisotropic velocity distribution, like
%that in disks, and consequently gives a better description of dynamical friction in disks. 
{%\bf
\item \label{enu:B}correct Chandrasekhar's formula to give improved predictions in peculiar situations.
}
\end{enumerate}{%\bf

Although approach \ref{enu:B} is theoretically more limited than approach \ref{enu:A},  for lack of fundamental insight and design for a given peculiar situation, it retains value in actionable power, as the path \ref{enu:A}, despite clarifying the limits of Chandrasekhar's formula with some improved formulas (e.g., Maoz 1993), did not find a complete solution to the dynamical friction problem.  

For example,  approach \ref{enu:B} can improve the formula to reduce its discrepancy with simulations prediction for timescale of spiraling of objects in a system with
cored dark matter halo of constant density distribution (e.g., Petts, Gualandris \& Read 2015), but cannot solve the problem of evaluation of dynamical friction in head-on encounters.

In this work we will adopt a type \ref{enu:B} approach and focus on the core stalling problem, i.e. the unability of Chandrasekhar's formula to predict the
stalling of infall of objects in cored systems.
In the last few years, several authors have related the problem to the DF timescale ($t_{df}$) of a GC orbiting dwarf galaxies or of infalling satellite galaxies in clusters (Read {\it et al.} 2006; Goerdt et al. 2006;  S\'anchez-Salsedo {\it et al.} 2006; Nath 2008; Cowsik {\it et al.} 2009; Inoue 2009; Namouni 2010, Gan {\it et al.}  2010). In particular, the DF effects for dwarf galaxies with cored dark matter halo of constant density distribution have been found to be considerably modified {%\bf 
(i.e. not experiencing dynamical friction (Goerdt 2010)).}  $N$-body simulations \citep{Goerdt2006, Inoue2009,Goerdt2010} %(Goerdt 2010)[$<-$INCLUDE WITH THE REST OF CITES]
%and semi-analytical models 
have shown that the sinking timescale of GCs to the galactic center may exceed the age of the universe, {%\bf 
such that they appear to stall at the edge of the core (Goerdt 2010).}

The reason for this stalling is interpreted in Goerdt (2006) as orbit-scattering resonance, or corotating state: perturber and background reach a stable state characterized by no angular momentum exchange. Inoue (2009) disagree with Goerdt (2006, 2010) on that interpretation.
In any case, all simulations agree with the stalling, contrary to Chandrasekhar's formula. Here we propose a solution based on a proper extension of the underlying statistical approach.

The so-called nonextensive statistical approach  provides an analytical extension of Boltzmann-Gibbs (BG) statistical mechanics very suitable to include effects of long-range forces and/or mildly out of thermal equilibrium states.} This ensemble theory is based on the formulation of a generalized entropy  proposed by Tsallis (1988,2009)

\begin{equation}
\label{Sq} S_{q} = k_{B}\frac{1-\sum_{i=1}^{W}p^{i}}{1-q},
\end{equation}
which reduces in the limit $q \rightarrow 1$ to the BG entropy $S_{BG} = -k_{B}\sum_{i=1}^{W}p^{i}\ln p_{i}$, since $p_{i}$ is the probability of finding the systems in the microstate $i$, $W$ is the number of microstates and $k_{B}$ is the Boltzmann constant. However, when the index $q \neq 1$, the entropy of the system is nonextensive, i.e, given two subsystems $A$ and $B$, the entropy is no more additive in the sense that $S_{q}(A+B) = S_{q}(A)+S_{q}(B)+(1-q)S_{q}(A)S_{q}(B)$.  The long-range interactions are associated to  the last term on the r.h.s. which accounts for correlations between the subsystems with the index $q$ quantifying the degree of statistical correlations.  Such a statistical description has been successfully applied to many complex physical systems ranging from physics to astrophysics and plasma physics, among which:  the electrostatic plane-wave propagation in a collisionless thermal plasma (Lima, Silva \& Santos 2000), the peculiar velocity function of galaxies clusters \citep{Lavetal98}, gravothermal instability \citep{TaSak2002}, the kinetic concept of Jeans gravitational instability (Lima, Silva \& Santos 2002),  and the radial and projected density profiles for two large classes of isothermal stellar systems \citep{LiSouza2005}. A wide range of physical applications can also be seen in Gell-Mann \& Tsallis 2004 (see also http://tsallis.cat.cbpf.br/biblio.htm for an updated bibliography).

In this paper, by assuming that a self-gravitating collisionless gas is described by the nonextensive kinetic theory (Silva {\it et al.} 1998; Lima {\it et al.} 2001), we  derive a new analytical formula for DF which generalizes the Chandrasekhar result. As an application, the DF timescale ($t_{df}$) for GCs falling in the galaxies center is  derived for the case of a singular isothermal sphere. This result suggest that the long timescales for GCs can be understood as a departure 
from the extensive regime. 
%
%In other words, there is no suppression of the DF since the long time can be just the statistical price to pay by the presence of long range forces %acting on the gravitational systems. 
%
   
%The paper is organized as follow: In section 2, we present a review of the basic Chandrasekhar formalism. In  section 3, we deriving an analytic formula for the DF in the context of %the nonextensive kinetic theory. In section 4, we utilize this new kind of the $q$-DF to calculate the dynamical timescale ($t_{fric}$) for a massive objects spirals to the center of %the host galaxy. Finally, in section 5 we presents our conclusions and some prospect for future works.

%\section{The Basic Chandrasekhar Approach}

\section{Dynamical Friction and Nonextensive Effects}

By following Chandrasekhar (1943), the DF deceleration on a test mass $M$ moving with velocity $v_{M}$ in a homogeneous and isotropic distribution of identical 
field particles of mass $m$ and number density $n_{0}$ reads:
\begin{equation}
\label{eq1}
\frac{d{\bf v_{M}}}{dt} = -16 \pi^{2} (\ln \Lambda) G^{2}Mm \frac{\int_{0}^{v_{M}}f(v)v^{2}dv}{v^{3}_{M}} {\bf v_{M}},
\end{equation}
where $G$ is the gravitational constant, $m$ is the mean mass of field stars and $f(v)$ represents their velocity 
distribution. The parameter $\Lambda = p_{max}/p_{min}$ depends on the ratio of the maximum ($p_{max}$) and minimum ($p_{min}$) impact parameters 
of the encounters contributing to generate the dragging force. 
%In the most of common situations it is considered that $p_{max} \sim L$ being $L$ the size of system and $p_{min} \sim l$ where $l$ is the size of object which can be a GC or black %hole. But in fact a sensitive determination of $\Lambda$ is a delicate problem. 

In the applications of DF, it is usually assumed that the distribution function of the stellar velocity field can be described by a Maxwellian distribution \citep{BT2008,Fellhauer2008}
\begin{equation}
\label{eq2}
f(X_{\star})=\frac{n_{0}}{(2\pi \sigma^{2})^{3/2}}e^{-X_{\star}^{2}},
\end{equation}
where $X_{\star} = v/\sqrt{2}\sigma$ denotes a normalized velocity with $\sigma$ indicating their dispersion.
The integration of (\ref{eq1}) results in:
\begin{eqnarray}
\label{eq3} \frac{d{\bf v_{M}}}{dt} = -\frac{4\pi \ln \Lambda G^{2} M \rho(r)} {v^{3}_{M}}H_{1}(X_{M})
{\bf v_{M}},
\end{eqnarray}
where $\rho(r) = n_{0}m$ and the function $H_{1}(X_{M})$ is given by
\begin{eqnarray}
\label{eq3a} H_{1}(X_{M}) = erf(X_{M})-\frac{2X_{M}}{\sqrt{\pi}}e^{-X^{2}_{M}},
\end{eqnarray}
with  $erf(X_{M})$ defining the error function as
\begin{equation}
\label{eq4}erf(X_{M})=\frac{2}{\sqrt{\pi}}\int_{0}^{X_{M}}e^{-X_{\star}^{2}}dX_{\star}.
\end{equation}

%\section{Dynamical Friction and Nonextensive Effects}

Now, in order to investigate the nonextensive effects on the Chandrasekhar theory, let us consider that the stellar field obeys the following power-law (Silva, Plastino \& Lima 1998, Lima, Silva \& Plastino 2001,  Lima \& de Souza 2005): 

\begin{equation}
\label{eq5} f(X_{\star}) = \frac{n_{0}}{(2 \pi \sigma^{2})^{3/2}}A_{q}e_q(X_{\star})
\end{equation}
where the so-called $q$-exponential is defined by
\begin{equation}
e_q(X_{\star}) = \left[1-(1-q)X_{\star}^{2}\right]^{\frac{1}{1-q}},
\end{equation}
and the quantity $A_q$ denotes a normalization constant which depends on the interval of the $q$-parameter. For values of $q < 1$, the positiveness of 
{the power argument means that the above distribution} exhibits a cut-off in the maximal allowed velocities. In this case, all velocities lie on the interval $(0,v_{max})$ and their maximum value is $v_{max} = \sqrt{2}{\sigma} / \sqrt{1-q}$. Taking this into account one may show that the normalization constant $A_q$ can be written in terms of Gamma functions as follows:
\begin{eqnarray}
\label{eq5b} 
%\left
{\begin{array}{ll}
A_{q} = (1-q)^{1/2}(\frac{5-3q}{2})(\frac{3-q}{2})
\frac{\Gamma(\frac{1}{1-q}+\frac{1}{2})}{\Gamma(\frac{1}{1-q})}, \,\,\,\hbox{$q < 1$} 
\\
A_{q} = (q-1)^{3/2} \frac{\Gamma(\frac{1}{q-1})}
{\Gamma(\frac{1}{q-1}-\frac{3}{2})}, \,\,\,\,\, \hbox{$q > 1$}
\end{array}}
%\right\}
\end{eqnarray}

\begin{figure*}
\centering
\includegraphics[width=90mm]{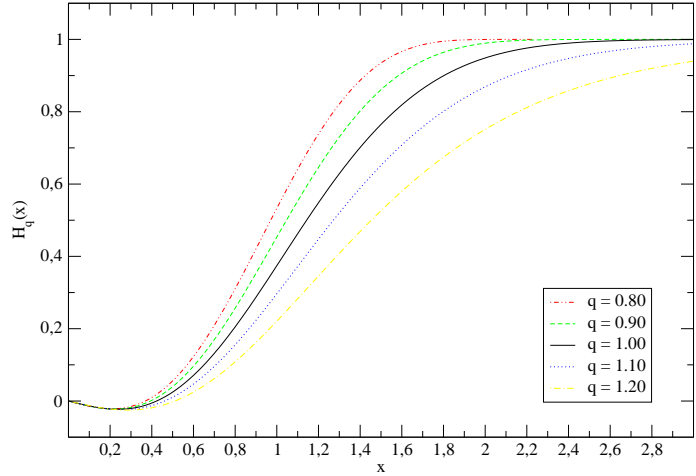} 
\caption[]{The $H_{q}(X)$ function. The solid black curve is the result based on Chandrasekhar theory ($H_{1}(x)$). The remaining  curves show the q-corrections for several values of the $q$-index. }
\end{figure*}

For generic values of $q \neq 1$, the DF (\ref{eq5}) is a power law, whereas for $q = 1$ it reduces to the standard Maxwell-Boltzmann distribution function (\ref{eq2}) since $A_{1} \rightarrow 1$ at this limit. Formally, this result follows directly from the known identity, $\rm{lim}_{d \rightarrow 0} (1 + dy)^{1 \over d} = {\rm{exp}(y)}$ \citep{Abr1972}.  The distribution (\ref{eq5}) is uniquely determined from two simple requirements (Silva {\it et al.} 1998): (i) isotropy of the velocity space, and (ii) a suitable nonextensive generalization of the Maxwell factorizability condition, or equivalently, the assumption that $f(v)\neq f(v_x)f(v_y)f(v_z)$. The kinetic foundations of the above distribution were also investigated in a deeper level through the generalized Boltzmann's equation. In particular, it was also shown that the kinetic version of the Tsallis entropy satisfies an extended $H_q$-theorem (Lima, Silva \& Plastino 2001).

Now, by considering that the power-law distribution (\ref{eq5}) is a valid description for the stellar velocity distribution we conclude that the expression describing the DF in this extended framework takes the following form:
\begin{eqnarray}
\label{eq6} \frac{d{\bf v_{M}}}{dt} &=& -\frac{16\pi^{2}G^{2} (\ln \Lambda) M\rho(r)}{v_{M}^{3}}A_{q} \times \nonumber\\&& \int^{X_{M}}_{0}X^{2}_{\star}e_{q}(X_{\star}) dX_{\star}\bf{v_{M}},
\end{eqnarray}
%The integral above is more easily solved by introducing the following relation
%\begin{eqnarray}
%\label{eq6a}X_{\star}e_{q}(X_{\star})dX_{\star}=-\frac{1}{2}e_{q}^{1-q}(X_{\star})d[e_{q}(X_{\star})].
%\end{eqnarray}
%By performing an integration by parts we obtain:
%\begin{eqnarray}
%&&\int^{X_{M}}_{0}X^{2}_{\star}e_{q}(X_{\star}) dX_{\star} = \nonumber\\&& \frac{1}{5-3q}\left\{\int^{X_{M}}_{0}e_{q}(X_{M})dX_{\star} -  X_{M}e_{q}^{2-q}(X_{M})\right\}. \nonumber\\&&
%\end{eqnarray}
which after an elementary integration can be rewrite as:
\begin{eqnarray}
\label{eq6b} \frac{d{\bf v_{M}}}{dt} = -\frac{4\pi G^{2}\ln \Lambda \rho(r)M}{v_{M}^{3}}\left(\frac{2}{5-3q}\right)H_{q}(X_{M})
\bf{v_{M}}, \nonumber\\&&
\end{eqnarray}
where $H_{q}(X_{M})$ is the general function depending on the $q$-parameter (compare with Eq. (5))

\begin{eqnarray}
\label{eq6c}H_{q}(X_{M}) = I_{q}(X_{M})- \frac{2X_{M}}{\sqrt{\pi}}A_{q}e_{q}^{2-q}(X_{M}).
\end{eqnarray}
In the above expression, the integral 
\begin{eqnarray}
\label{eq6d}I_{q}(X_{M}) = \frac{2A_{q}}{\sqrt{\pi}}\int_{0}^{X_{M}}e_{q}(X_{\star})dX_{\star},
\end{eqnarray}
is the $q$-generalization of the error function (see Eq.(6)).
Fig. 1 plots $H_{q}(X)$ for $q$ in the range 0.80, 1.20.  

As one may check, the nonextensive expression for the DF (including the auxiliary functions $H_q$ and $I_q$) reduces to the Chandrasekhar result in the Gaussian limit ($q \rightarrow 1$). {This } shows clearly that the  collective effect from gravitational interactions of $M$ (with all stars of the field) is strongly dependent on the statistical model.  An interesting aspect of the above formulae is that the {results are given by  analytical expressions}. In principle, they can be useful for semi-analytical implementations because the easy comparison with the standard approach (see next section). Naturally, we are also advocating here that the idealized framework based on the Maxwellian distribution (Chandrasekhar 1943) may be in the root of some theoretical difficulties shown by $N$-body simulations, like the ones related to the decay orbits of GCs.

\section{Decay of Globular Orbits}

In order to illustrate some consequences of the above derivation, let us now analyze the  nonextensive solution for the decaying orbit of a GC in the stellar galactic field. 
%This is an instructive problem first because it constitutes an intensively studied subject, and, also because the basic results can be expressed by analytical expressions. 
As a GC orbits through the galactic field, it is subject to DF due to its interaction with the stellar distribution.  By assuming spherically symmetric star distribution, the dragging force  decelerates the cluster motion which loses energy thereby spiraling toward the galaxy center. Therefore, whether the GC is initially on a circular orbit of radius $r_{i}$, it is convenient to define an average DF timescale, $t_{df}$, as the time required for the cluster reach the galaxy center. For simplicity's sake, we also consider that the mass density distribution of the galaxy is described by the singular isothermal sphere 
\begin{equation}
\label{eq7}\rho(r)=\frac{1}{4\pi G}\left(\frac{v_{c}}{r}\right)^{2},
\end{equation}
with $v_{c}$ being circular speed and $\sigma = v_{c}/\sqrt{2}$ the velocity dispersion. This simplified mass distribution has the benefit of having a planar rotation curve and therefore might be considered as a crude but minimally realistic distribution for the external region of normal galaxies.
{The frictional force felt by a cluster of mass $M$ } moving with speed $v_{c}$ through the stellar field now reads:
\begin{eqnarray}
\label{eq8}F = -\left(\frac{2}{5-3q}\right) G \ln\Lambda \left(\frac{M}{r}\right)^{2} H_{q}(1),
\end{eqnarray}
where $H_{q}(1)$ is the general function (\ref{eq6c}) written in the coordinate $X = (v_{c}/\sigma \sqrt{2}) = 1$. Note also that the integral $I_{q}(X_{M})$ defined in (\ref{eq6d}) now reduces to
\begin{eqnarray}
I_{q}(1) = \frac{2A_{q}}{\sqrt{\pi}} {_{2}F_{1}} \left(\frac{1}{q-1}, \frac{1}{2}; \frac{3}{2}; 1-q \right),
\end{eqnarray}
where ${ _{2}F_{1}}(a, b; c; z)$ is the Gauss hypergeometric function. Either from the  above representation or from the  integral form (\ref{eq6d}), we see that the error function $erf(1)$ is obtained as a particular case in the extensive regime, that is,  $I_{1}(1) = erf(1) \approx 0.8427$ \citep{Abr1972}. It means that $H_{1}(1) = erf(1)-(2/\sqrt{\pi})e^{-1} \approx 0.428$ \citep{BT2008} .

Now, returning to expression (\ref{eq8}), we recall that the dragging force is tangential to the cluster orbits, and, therefore, the cluster  gradually loses angular moment per unit mass $L$ at a rate $dL/dt = Fr/M$. Since $L = rv_{c}$ we can rewritten equation (\ref{eq8}) as
\begin{eqnarray}
\label{eq8b}r\frac{dr}{dt}=-\left(\frac{2}{5-3q}\right) \left(\frac{GM}{v_{c}}\right)\ln\Lambda H_{q}(1).
\end{eqnarray}
By solving this differential equation subjected to the initial condition, $r(0) = r_{i}$, we find that the cluster {reach the galactic} center after a time
\begin{eqnarray}
\label{eq8c}t^{(q)}_{df} &=& \left(\frac{5-3q}{2}\right)\frac{0.5v_{c}r_{i}^{2}}{GM \ln\Lambda H_{q}(1)}.
\end{eqnarray}
This nonextensive timescale for decaying orbits of GCs generalizes  the Chandrasekhar result (see Binney \& Tremaine 2008) which is readily recovered in the Gaussian extensive limit ($q=1$).

\begin{figure*}
\centering
%\hspace{-0.5cm}
\includegraphics[width=180mm]{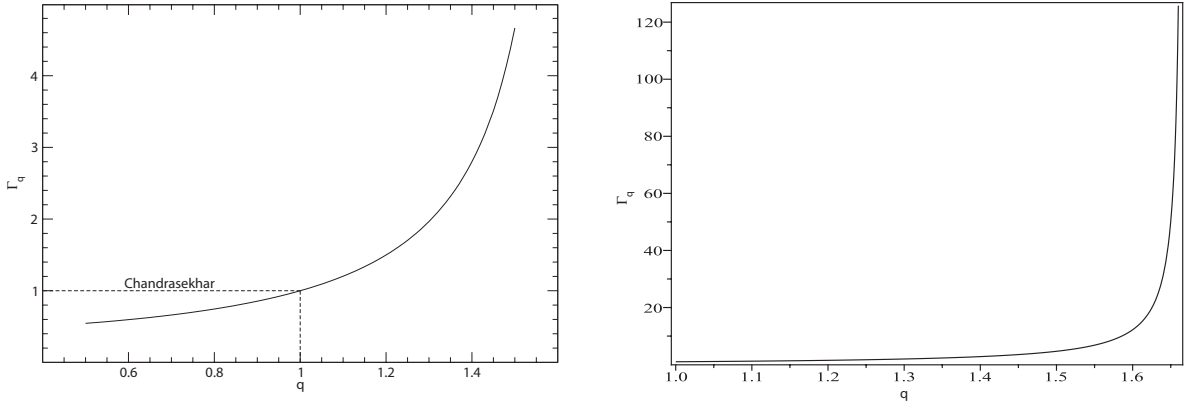} 
\caption[]{Behavior of the relative time scale ratio  $\Gamma_{q}$. {%\bf 
Left panel}: we see that for $ q > 1$, the characteristic nonextensive time scale for dynamic friction can be much greater than in the standard Chandrasekhar approach. The right panel shows $\Gamma_{q}$ only for $q>1$, and for larger values of panel a.
}
\end{figure*}

At this point, it is interesting to compare the above nonextensive prediction with the standard result based on the Chandrasekhar approach. To begin with, let us  assume typical values for the parameters $r_{i}$, $v_{c}$ and $M$, namely: $r_{i} = 2Kpc$, $v_{c} = 250kms^{-1}$ and 
$M = 10^{6}M_{\odot}$.  With these choices we get:
\begin{eqnarray}
t^{(q)}_{df} & \approx & \frac{1.14 \times 10^{11}}{H_{q}(1) \ln\Lambda} \left(\frac{5-3q}{2}\right)\left(\frac{r_{i}}{2kpc}\right)^{2} \nonumber\\&& \times \left(\frac{v_{c}}{250kms^{-1}}\right) \left(\frac{10^{6}M_{\odot}}{M}\right)yr,
\end{eqnarray}
which reduces to the standard value in limiting case ($q = 1$) as given by Binney \& Tremaine (2008). The nonextensive 
corrections are more directly  quantified by introducing the dynamic time ratio, $\Gamma(q) \equiv t^{(q)}_{df}/t^{(1)}_{df}$, where  $t^{(1)}_{df}$ denotes the Chandrasekhar result. By using (18) we find
\begin{eqnarray}
\label{eq8d}\Gamma(q) = \left(\frac{5-3q}{2}\right)\frac{H_{1}(1)}{H_{q}(1)}=\left(\frac{5-3q}{2}\right)\frac{0.428}{H_{q}(1)}.
\end{eqnarray}
where the function $H_q(X)$ was defined by Eq. (11). 

In Figure 2, {%\bf 
left panel}, we display the nonextensive corrections for a large range of the nonextensive $q$-parameter. As a general result, we see that the $\Gamma(q)$ ratio is strongly dependent on the q-parameter. The nonextensive time scale is greater or less than the extensive Chandrasekhar result depending on the interval of the $q$-parameter. Note also that $t^{(q)}_{df}$ is greater or smaller than $t^{(1)}_{df}$ if $q > 1$ or $q < 1$, respectively. 

{%\bf 
The key point, explaining the stall showed in Fig. 3, lies in the right panel of Fig. 2: the fast steepening of the $\Gamma(q)$ ratio with $q$,  showing the infalling time quickly becomes very large with growing $q$. For $q=1.66$, 
$\Gamma(q) \simeq 126$ while for $q=1.6666666$, $\Gamma(q) \simeq 1.26 \times 10^7$.  }

\begin{figure*}
\centering
~~\hspace{02.5cm}\includegraphics[width=180mm]{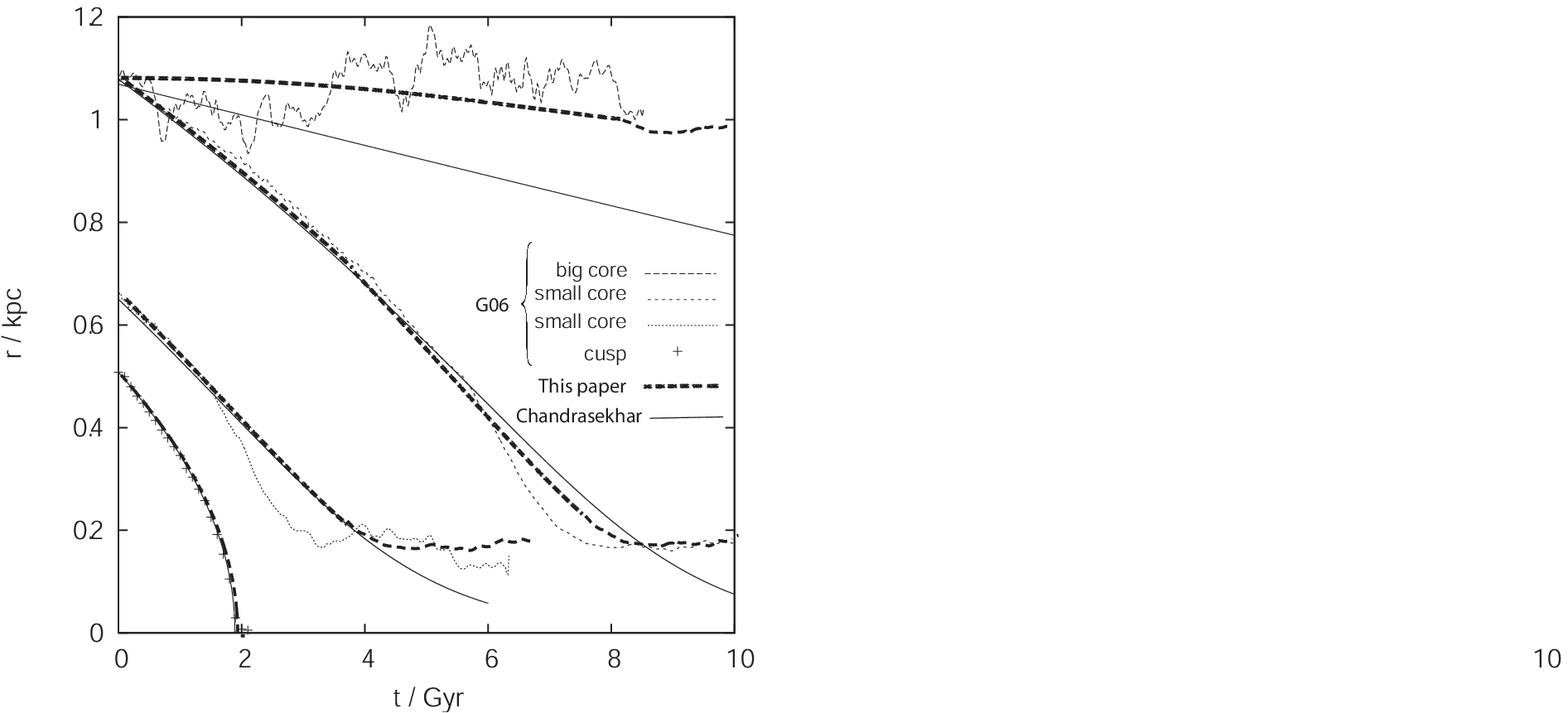}
\caption[]{\label{TimeEvol} Time evolution of the radial distance of a single globular cluster from the center of the host. The solid line represents the case of {%\bf% 
standard} Chandrasekhar's formula ($q=1$), the thick dotted line is the result of this paper. The thin dashed line, the dotted lines, and the {%\bf% 
plusses} represent the G06 results for the big core, the small core, and the cusp case respectively.    
%and the curve in the top the results of G06 simulations for a cored profile of the host. The dotted line represent the time evolution of the radial distance for the modified Chandrasekhar's formula for the case $q=1.2$.
}
\end{figure*}

%\begin{figure}[ptbh]
%\centerline{\epsfysize=45mm\epsffile{goerdt2_1.eps}} 
%\caption{Time evolution of the radial distance of a single globular cluster from the center of the host. Left panel: the solid %line represents the case of Chandrasekhar's formula ($q=1$), and the crosses the results of G06 simulations for a cuspy profile %of the host. The dotted, dashed, dot-dashed lines represent the time evolution of the radial distance for $q=1.2$, 1.4, 1.5. %Righ panel: similar to the left one but for a cored proile of the host. The solid line is the case of Chandrasekhar's formula %($q=1$), while the dotted line the case $q=1.2$. The top curve represents G06 simulations.  } 
%\end{figure}

%\subsection{The instantaneous orbital decay}
%It is also widely known that the orbital velocity $v_{c}$ is a function of $r$. Therefore, it is natural to ask about the instantaneous orbital decaying process.
%In order to answer this question we consider  that the angular momentum per unit mass is now given by $L = rv_c(r)$. In this case, Eq. (16) take the following form:  
%\begin{eqnarray}
%\label{eq9}\frac{dr}{dt} = -\frac{4\pi G^{2}M\ln \Lambda }{v_{c}^{2}\frac{d}{dr}[rv_{c}(r)]}\left(\frac{2}{5-3q}\right) r \rho(r)H_{q}(X_{c}(r)),\nonumber\\&&
%\end{eqnarray}
%where $X_{c}(r) = v_{c}(r)/\sigma(r)\sqrt{2}$ and the q-function $H_{q}(X_{c}(r))$ which depends on $v_{c}(r)$ and $\sigma(r)$,  is  defined by% 

%\begin{eqnarray}
%\label{eq10}H_{q}(X_{c}) = I_{q}(X_{c})-\frac{2X_{c}(r)}{\sqrt{\pi}}A_{q}e_{q}(X_{c}(r)).
%\end{eqnarray} 

\begin{figure*}
\centering
%\hspace{-0.5cm}
\includegraphics[width=100mm]{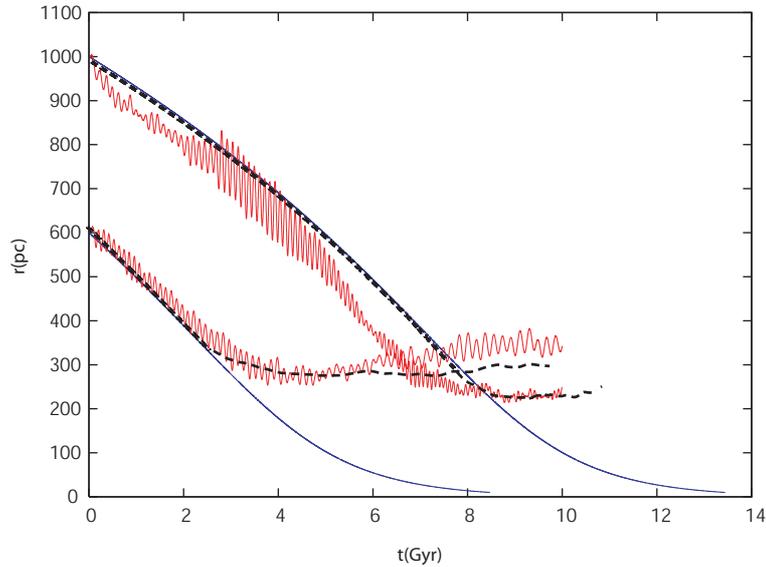} 
\caption[]{Time evolution of the radial distance of a single globular cluster from the center of the host. The solid line represents the case of Chandrasekhar's formula ($q=1$), the thick dotted line is the result of this paper. The waving jagged lines are the result of Inoue (2009) simulations. 
%and the curve in the top the results of G06 simulations for a cored profile of the host. The dotted line represent the time evolution of the radial distance for the modified Chandrasekhar's formula for the case $q=1.2$.
}
\end{figure*}

\subsection{Comparison with N-body simulations 
%Goerdt's result 
}

{

In the introduction, we already discussed the globular clusters (GCs) puzzle.
%, namely the too short sinking time of GCs to the dwarf galactic center if the density profile of the dwarf is cuspy. 
%As discussed, 
{%\bf 
Several solutions have been proposed, one based on the presence of a cored inner density profile in dwarf galaxies, and the shortcomings of Chandrasekhar's formula (Goerdt et al. 2006 (G06)). While simulations show a drastic reduction of dynamical friction in cores, leading infalling objects to stall, Chandrasekhar formula does fail to reproduce it.}

%As already discussed, in dwarf galaxies (e.g., Fornax), the sinking time of globular clusters (GCs) to the center of dwarf %galaxies is much smaller than the age of the universe, while in some dwarf galaxies GCs are still present.  
%the globular clusters puzzle in dwarf galaxies like Fornax, is the 

%
%Here, we follow their analysis, substituting the classical Chandrasekhar?s formula with the modified formula we got
%
 
In the following, we see {%\bf 
how our modified formula, contrary to the classical Chandrasekhar's, can reproduce G06's and  Inoue (2009) results.}

Similarly to G06, we use the generalized density profile depending on $\alpha, \beta,
\gamma$ (Hernquist 1990):
\begin{equation}
\rho(r)=\frac{\rho_0} {\left({\frac{r}{r_{\rm s}}}\right)^\gamma \left[{1 +
\left({\frac{r}{r_{\rm s}}}\right)^{\alpha}}\right]^{\frac{\beta - \gamma}
{\alpha}}}
\label{eq:prof}
\end{equation}

We calculated the radial distance of the GC from the host center using 
%the cuspy and a 
the "big cored" profiles, as defined in G06: $\rho_0=0.10 M_{\odot}/pc^3$, $r_s=2.2$ kpc, $M_{\rm vir}= 3.0 \times 10^{10} M_{\odot}$, $\gamma=0$, $\beta=3$, $\alpha=1.5$ (see their Fig.~1). {%\bf 
The "small cored" profile used the same parameters as for the "big cored" profile except for $r_s=0.91$ kpc, and $M_{\rm vir}= 2.0 \times 10^{9} M_{\odot}$. The cuspy profile had $\rho_0=0.0058 M_{\odot}/pc^3$, $r_s=2.4$ kpc, $M_{\rm vir}= 2.0 \times 10^{9} M_{\odot}$, $\gamma=1.5$, $\beta=3$, and $\alpha=1.5$ .}

Instead of using Chandrasekhar's formula and Eq. 6 of G06, we used the modified version of Chandrasekhar's equation, which led us to write 
\begin{equation}
{dr \over dt} =  -{4 \pi {\rm ln} \Lambda (r) \rho(r) G^2 M_{\rm GC} r \over
v^2_{\rm c}(r) {d(rv_{\rm c}(r)) \over dr}} 
\left(\frac{2}{3-5q}\right) H_q
\label{eq:new}
\end{equation}
where $\rho(r)$ is given by Eq. (\ref{eq:prof}), the Coulomb logarithm ${\rm ln} \Lambda (r)$ is given by
\begin{equation}
{\rm ln} \Lambda (r) ={\rm ln} {b_{\rm max} \sigma^2(r) \over GM_{\rm GC}},
\end{equation}
being $b_{\rm max}$ is the largest impact parameter, and which is put equal to $b_{\rm max}=0.25$ kpc for the cuspy profile, and 
$b_{\rm max}=1.0$ kpc for the cored one (similarly to G06), $v_{\rm c}(r)$ is the circular velocity, defined by 
$v_{\rm c}=\sqrt{r d\Phi(r)/dr}$, where $\Phi$ is the gravitational potential corresponding to Eq. (\ref{eq:prof}). 
Finally, the velocity dispersion $\sigma(r)$ is given by Eq. 3 in G06, namely
\begin{equation}
\sigma^2(r) = {1 \over \rho(r)}\int^{\infty}_{r}{{M(r') \rho(r')\over r'^2}
dr'}.
\end{equation}

%
%In Fig. 3, we plot the time evolution of the radial distance. The solid line, in the left panel of Fig. 3 show the $r(t)-t$ %relation when using the Chandrasekhar's formula, while the crosses are the results of the G06 simulation. Simulations and %analytical predictions are in good agreement    
%
{%\bf
Fig. 3 displays the comparison of our, and Chandrasekhar's, formula for a single globular cluster's radial evolution towards its host's center in G06 simulations. 
% of a cored profile. 
The solid lines represent the results of {%\bf%
standard Chandrasekhar's formula (using Maxwell's PDF, i.e. $q=1$)}, while the thick dotted lines, the results of this paper, displayed in  each case accompanied by the corresponding G06 results, marked by the thin dashed line, the dotted lines, and the {%\bf%
plusses}, for the big core, the small cores, and the cusp case respectively. 

The Chandrasekhar's formula correctly predicts the cuspy profile, while it overestimates continued sinking towards the center for all other cases, in disagreement with G06 simulations. The latters show an initial Chandrasekhar sinking, an accelerated sinking of GC  (super-Chandrasekhar phase) at core  entry, followed by a rapid stalling.  

Our approach is equal to Chandrasekhar's  and $q=1$ for the cuspy profile. For the small and big cored profiles, $q=1.66$, and a stalling is predicted. Similarly to Petts, Gualandris \& Read (2015) (see the following), our model doesn't exhibit super-Chandrasekhar phase. 

Similar comparisons to Inoue (2009) simulations are made in Fig. 4. They followed Read et al (2006), and G06, using the Hernquist's profile (Eq. \ref{eq:prof}), with $\rho_0= 0.10 M_{\odot}/pc^3$, scale radius $r_s=0.91$ kpc, an almost constant density within 200-300 pc, and a velocity dispersion obtained from Jeans equation, as in G06. A single GC, on initial circular orbit radius 600 pc or 1 kpc, is followed. 

The same line conventions are used to represent this paper's results and those of Chandrasekhar's formula ($q=1$). Inoue (2009)'s simulations yield the wavy jagged lines.
As in Fig. 3, Chandrasekhar's formula does not show any stalling, while our model does, but fails, similarly to Petts, Gualandris \& Read (2015), to reproduce the farthest initial GC super-Chandrasekhar's phase. 

From these comparisons, we can conclude that the extra parameter $q$ models the effects of non-locality expected on DF beyond Chandrasekhar's formula: Petts, Gualandris \& Read (2015) obtained  core stalling assuming Coulomb logarithm radial dependence, with a null DF for trajectories with impact parameter smaller or equal to a minimum , $b_{\rm min}$, representing absence of particles to scatter off the satellite. They conjectured that the super-Chandrasekhar phase could be reproduced by taking into account either super-resonance in the core (e.g., Goerdt 2010), or faster than satellite host's stars. The fundamental importance of the latter was confirmed by Antonini \& Merritt (2012) in black hole inspiral with low density background. Their results are further improved using a time dependent distribution function directly extracted from simulations.
The principal drawback of such improvement is the need to run a simulation to obtain the time dependent distribution, before inputting it in the 
Antonini \& Merritt (2012) approach. This issue obviously reduces the predictive power of that method, since the simulation already contains the correct description of the motion of the infalling body.  This problem is not present in Petts, Gualandris \& Read (2015) or in the presently proposed method.
However, the result of Antonini \& Merritt (2012) is a further confirmation that one limit of Chandrasekhar’s formula lies in its assumption of locality (two body interactions).
The non-locality of beyond Chandrasekhar DF also agrees with results from approach \ref{enu:A}: Maoz (1993) (see also Del Popolo \& Gambera 1999; Del Popolo 2003) showed DF depends on the global structure of the system, given by a new parameter ($q$ in the present work), and the DF force direction departs from the opposite to the particle's motion. 

Our present study finds $q$  adopts different values for cuspy and cored profiles to get correct $r-t$ behaviour. 

{%\bf 
At this stage, the question arises whether our semi-analytical model's stalling capture arises from the extra $q$, considered as a parameter chosen to reproduce simulations, or as realistic physics provided by the model.
To test this point, we use Eddington's formula to obtain the phase-space density function (PDF) $f(E)$ corresponding to the spherically symmetric density profiles (Eq. \ref{eq:prof}) we used.
Comparing the $f(E)$ obtained from the cuspy and cored profiles with our distribution function (Eq. \ref{eq5}) can give some hints on this issue, as
an agreement between the $f(E)$s obtained in each case from both methods, Eddington's or the choice of $q$ in Eq. (\ref{eq5}) could be considered to imply that our model contains realistic physics

Such a test is performed in Fig. \ref{EddingtonTest}.

\begin{figure*}
\centering
\includegraphics[width=80mm]{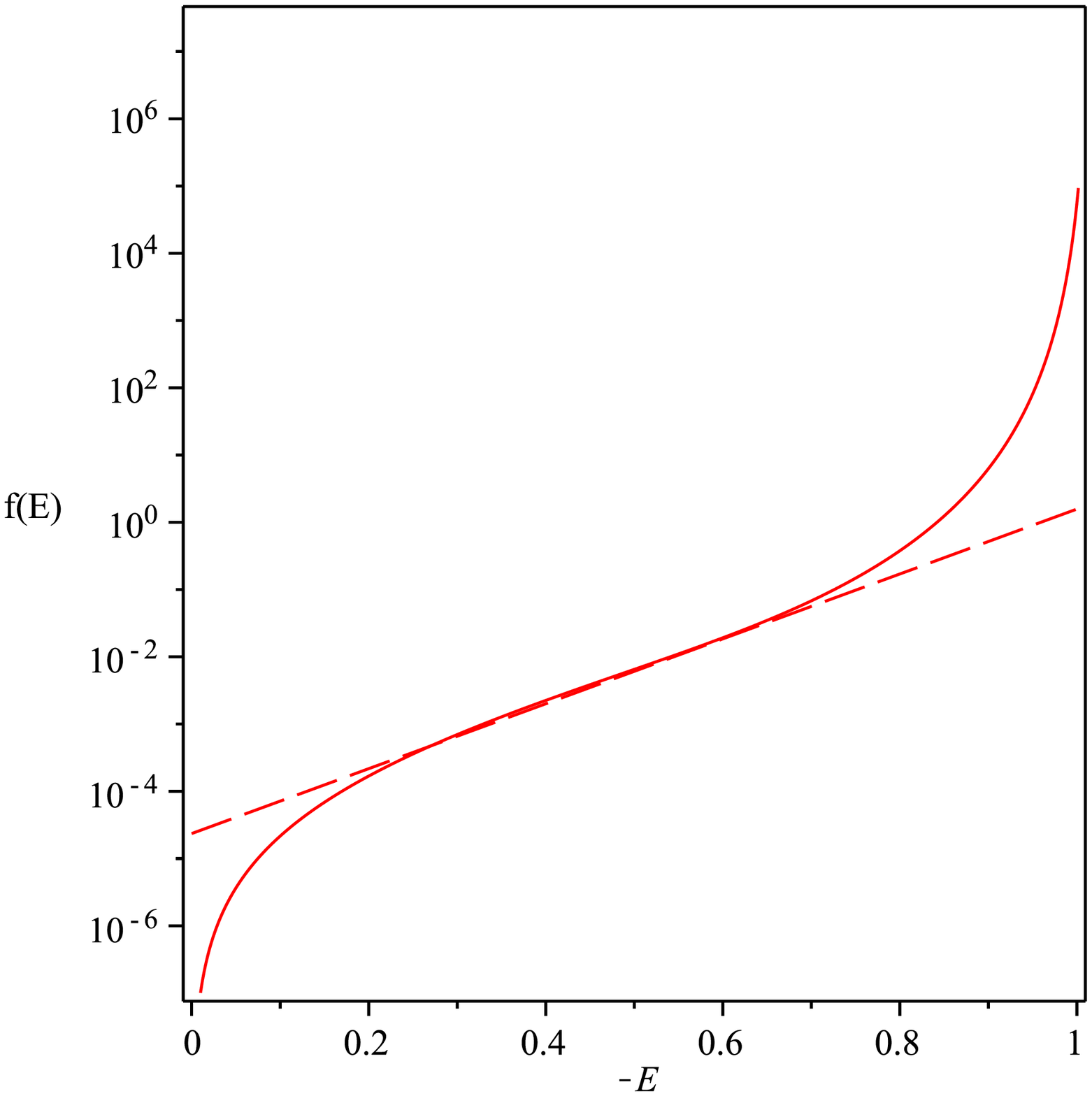}\includegraphics[width=80mm]{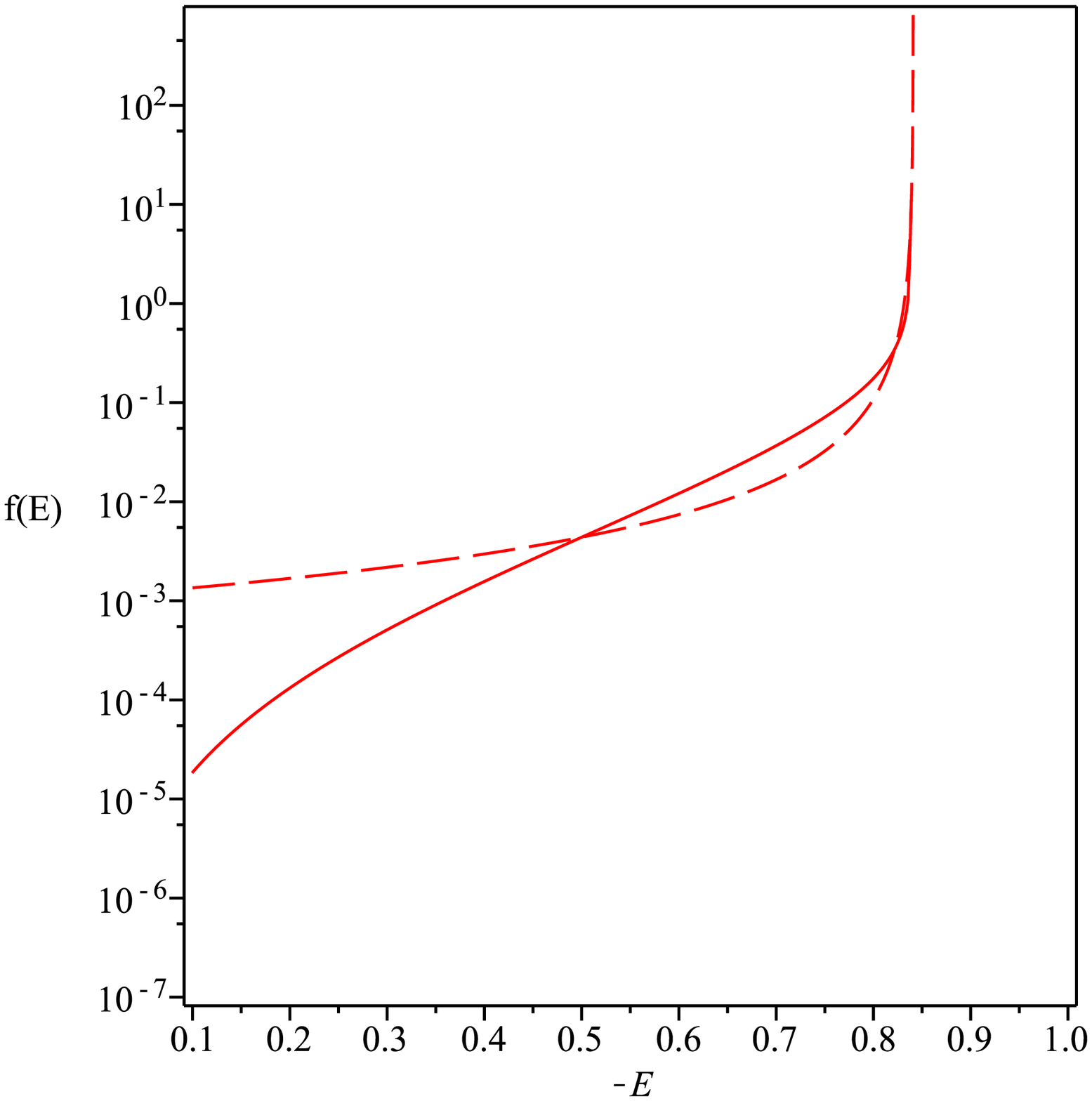}

\caption[]{%\bf 
\label{EddingtonTest}Comparison of the Eddington PDFs ($f(E)$, solid lines), for the cuspy (left panel) and small cored (right panel) profiles (eq. \ref{eq:prof}) respectively, with the corresponding Tsallis PDFs ($f(E)$, dashed lines, Eq. \ref{eq5}), that is for $q=1$ and $q=1.66$ respectively.
}
\end{figure*}

In the left panel of Fig. \ref{EddingtonTest}, the cuspy profile's PDF from Eddington's formula is represented by a solid line, while the dashed line displays the Tsallis' $f(E)$, obtained from Eq. (\ref{eq5}) for $q=1$. The right panel shows again Eddington's PDF as the solid line, this time for the small-cored profile, and that from Tsallis' (Eq. \ref{eq5}) with $q=1.66$ as the dashed line.

Except in a limited range of binding energies, the two curves are different. At face value, such result could be considered as an evidence that our model is more a useful ansatz to determine the $r-t$ evolution than a realistic physical model. A deeper insight leads us to a different conclusion.  

As is well known, and shown in Fig. \ref{TimeEvol}, the Chandrasekhar formula works very well for the cuspy profile. We also know that the underlying distribution of the Chandrasekhar formula is the Maxwell distribution, which coincides with the Tsallis statistics for $q=1$. On the base of the very good agreement between the Chandrasekhar formula and simulations in the case of cuspy profiles, one should expect a similarly good agreement of the cuspy profile PDF obtained from Eddington's formula with the Tsallis $f(E)$ for $q=1$. This is not the case, as shown by the left panel of Fig. \ref{EddingtonTest}!{%\bf%

Here, we should point out that the PDF obtained from Eddington's formula assumes there is no other physically meaningful parameter than binding energy $E$
  that determines the state of the spherical system. However, the density profiles (\ref{eq:prof}) are not provided with such assumption: they are issued from observation or Nbody simulations that involve all physical effects. Using the Eddington formula in such cases is biasing the obtained PDF.}
  
We must conclude that the distribution underlying Chandrasekhar's formula is only one of the keys of success of the Chandrasekar's formula, or Chandrasekar's-like formulas and that the success of the formula does not reflect exactly its precise PDF shape. Other physical factors are compounded in it that are equally important. Which is why our approach still retains physical meaning: physically, the correlations of the values of $q$ with cuspiness of the profiles reflect
that two body interactions dominate in steep density profiles whereas the
global structure of the halo is needed in cored profiles to fully
account for the effects of DF, as it is expected that long-range interaction
will make a system nonextensive \cite{Tsallis88} {%\bf%
(the PDF should be of the form $f\left(E,q\right)$
 , with $q$
  reflecting the importance of non-locality in the system)}}. We plan to extend that exploration, following Read et al. (2006), finding the infall dependence on density profile inner slope and satellite mass. Finally, a future study should explore if a  $q(r)$ dependence can reproduce the super-Chandrasekhar phase. 

}

%
%Altre verifiche future: dipendenza dalla slope e dalla massa come Read et al. 2006 e come q dipende dalla slope.
%
  
}
%shows a slowing down and then aof the infall when the GC moves towards the center containing less DM than the outer parts of the %host.

%The solid line shows the $r(t)-t$ relation when using the Chandrasekhar's formula, while the top curve the results of the G06 simulation. The dotted line represent the Chandrasekhar's modified formula for the case $q=1.2$. As the plot shows, Chandrasekhar's formula predicts a continued sinking towards the center, while the G06 simulations after an initial sinking shows a slowing down of the infall when the GC moves towards the center containing less DM than the outer parts of the host. This result is in good agreement with the radial infall found using the modified Chandrasekhar's formula ($q=1.2$) (dotted line). This confirms that apart the solutions proposed to solve the decay time puzzle of GCs in some dwarf galaxies can be solved in the framework of a $q$-non-extensive distribution. 

We want now to recall that the GCs decay time puzzle is strictly connected to another fundamental problem of the $\Lambda$CDM model, namely the nature of the inner density profiles of dwarf galaxies, the so called cusp/core problem (Flores \& Primack 1994; Moore 1994). The fact that the MW dSphs are not nucleated, and only around 30\% of those in clusters are, may imply that the inner density profiles of the MW dSphs are cored. Kleyna et al. (2003) studying the stellar number density in the Ursa Minor dSph (Umi dSph), concluded that the second peak in the stellar number density is unreconcilable with a cuspy profile. More recently many authors have studied the problem of the inner structure of the MW dSphs, since this could give information on the nature of DM. As discussed by several authors (e.g., Pe\~narrubia et al. 2012) cored inner profiles in low mass dSphs would increase the known tension between some of 
the small scale problems of the $\Lambda$CDM.  
Unfortunately, controversial evidences have been presented (e.g., Jardel et al. 2013; Amorisco \& Evans 2012; Jardel et al. 2013) to date due to     
the difficulty to distinguish cusps from cores (e.g., Strigari et al. 2014) and this leaves the inner structure of dSph galaxies as a still open debate.

\section{Conclusions}

We have derived the $q$-dynamic friction force for a point mass moving through a homogeneous background in the context of the nonextensive kinetic theory. Simple and analytical forms were obtained, and, as should be expected, they smoothly reduce to the standard Chandrasekhar results in the extensive limiting case ($q = 1$). However, for $q \neq 1$ a large variety of qualitatively different behaviors are predicted when the free parameter $q$ is continuously varied (see Figs. 1 e 2).  As an application, we have discussed 
{%\bf  
\renewcommand{\theenumi}{\alph{enumi}}\begin{enumerate} 
\item the dynamical timescale for a globular cluster collapsing to the center of a massive dark matter halo described by an isothermal sphere,  
\item we showed that the evolution of the radial distance in the case of the "small" and "big cored" profiles studied by Goerdt et al (2006) may be reobtained in the case $q=1.66$, and 
\item we confronted the $q$-modified Chandrasekhar formula with two sets of Nbody simulations that reveal how successful the model is at reproducing the core stalling. However it reveals that a super-Chandrasekhar phase is yet to be properly  modeled.\end{enumerate}}
 The results presented here suggest that the problem related to the large timescale shown by numerical $N$-body simulations and semi-analytical models may  naturally be solved (with no ad hoc mechanism) by taking a proper $q$-nonextensive distribution with parameter greater than unity. Applications to other
% more realistic 
density profiles like the lowered nonextensive halos distribution (Silva, de Souza \& Lima 2009; Cardone, Leubner \& Del Popolo 2011) and a detailed comparison with semi-analytical models will be discussed in a forthcoming communication. 
\vspace{0.2cm}

\noindent {\bf Acknowledgments:} JMS is supported by FAPESP Agency and JASL by FAPESP and CNPq (Brazilian Research Agencies).
The work of M.Le~D. has been supported by PNPD/CAPES20132029. M.Le~D. also wishes to acknowledge IFT/UNESP. ADP wishes to thank Dian-Yong Chen  from Laznhou IMP-CAS.

\end{document}